\documentclass[10pt,a4paper]{article}
\usepackage{amsmath}
\usepackage{amssymb}
\usepackage{amsfonts}
\usepackage{amstext}
\usepackage{amsbsy}
\usepackage[mathscr]{eucal}
\usepackage{graphicx}
\usepackage{color}
\usepackage[all]{xy}

\newcommand{\beq}{\begin{equation}}
\newcommand{\eeq}{\end{equation}}
\newcommand{\beqa}{\begin{eqnarray}}
\newcommand{\eeqa}{\end{eqnarray}}
\newcommand{\ket}[1]{| #1 \rangle}
\newcommand{\bra}[1]{\langle #1 |}

 \numberwithin{equation}{subsection}

\makeindex
\title{\Large\textbf{Toric varieties: Simple combinatorial and geometrical structure of multipartite quantum systems}}

\author{\textit{Hoshang Heydari}\\
        \small\textit{Institute of Quantum
Science, Nihon University,}\\
\small\textit{1-8 Kanda-Surugadai, Chiyoda-ku, Tokyo 101-8308,
Japan}
\\\small\textit{Email: hoshang@edu.cst.nihon-u.ac.jp}}

\date{}
\begin{document}
\maketitle \thispagestyle{empty} \maketitle 

\begin{abstract}
We investigate the geometrical structure of multipartite states based on the construction of  toric varieties. We show that the toric variety represents the space of  general pure states and projective toric variety  defines the space of separable set of  multi-qubits states. We also discuss in details the construction of single-, two-, three-, and multi- qubits states. This  construction  gives a very simple and elegant visual representation   of the geometrical structure of multipartite quantum systems.
\end{abstract}
\section{Introduction}

In recent years we have witnessed an increase of interest in geometrical and topological structure of bipartite and multipartite states \cite{Dorje99,Moss,Beng,Bern,Miyake,Levay,Grab}. We have also discussed geometrical and topological structure of multipartite states based on the Segre variety and Hopf fibration in
\cite{Hosh1,Hosh2}. In this paper we will give a simple construction of multi-qubit states based on toric varieties and projective toric varieties. Toric varieties provide an elementary way to see many examples and phenomena in algebraic geometry.
During recent decade toric varieties have been constructed in different contexts in mathematics \cite{Ewald,GKZ,Fulton,Oda} and more recently also they have found application in string theory in connection with the construction of mirror symmetric \cite{Vafa,Cox}. A toric variety $\mathbb{X}$ is a complex variety that contains an algebraic torus $T=(\mathbb{C}\backslash\{0\})^{n}$ as a dense open set and with action of $T$ on $\mathbb{X}$  whose restriction to $T\subset\mathbb{X}$ is the usual multiplication on $T$. There are two standard ways of characterizing toric varieties. Projective toric varieties can  be described by lattice point in a polytope and normal toric varieties can be   described by fan. Moreover, if the toric varieties are both projective and normal, then they can be described by either polytope or fan. In string theory, reinterpretation of certain data for a fan as data for polytope leads to geometric construction of mirror manifolds \cite{Vafa}. One of the interesting properties of toric varieties is that it is easy to describe $T$-invariant subvarieties. Let $\mathbb{X}$ be a projective variety and  $\mathcal{G}$ an algebraic group acting on $\mathbb{X}$. Then, in geometric invariant theory quotient of a projective variety $\mathbb{X}$ by $\mathcal{G}$  is also a projective variety. A toric variety can be defined to be a normal variety that admits a torus action with a dense orbit. In this paper we will show that the toric varieties and projective toric varieties can be used to give a simple   geometrical construction of multipartite systems. However this construction is heavily   based on new mathematical definitions and facts from combinatorial geometry and algebraic geometry which we will cover in sections \ref{cpv} and \ref{toric}. In section \ref{tvqs} we will describe the geometrical structure of multipartite quantum systems based on these toric varieties.

\section{Complex multi-projective variety} \label{cpv}
 In this section we will give a short introduction to the complex algebraic affine and projective varieties. We also discuss the construction of the multi-projective Segre variety.
 Interested reader may find more on
complex projective algebraic geometry in \cite{Griff78,Mum76,Hart77}.
\subsection{Projective algebraic variety }\label{pav}
Algebraic affine and projective varieties are the most fundamental structures of algebraic geometry. In this subsection we will give definitions and the basic properties of complex algebraic affine and projective  varieties and ideals.
Let $\mathbb{C}$ be a complex algebraic field. Then an affine
$n$-space over $\mathbb{C}$ denoted $\mathbb{C}^{n}$ is the set of
all $n$-tuples of elements of $\mathbb{C}$. An element
$P\in\mathbb{C}^{n}$ is called a point of $\mathbb{C}^{n}$ and if
$P=(a_{1},a_{2},\ldots,a_{n})$ with $a_{i}\in\mathbb{C}$, then
$a_{i}$ is called the coordinates of $P$.
 Let $\mathbb{C}[z]=\mathbb{C}[z_{1},z_{2}, \ldots,z_{n}]$ denotes the polynomial
algebra in $n$  variables with complex coefficients. Then a Zariski closed set in $\mathbb{C}^{n}$ is a set of common zeros of a finite number of polynomials from $\mathbb{C}[z]=\mathbb{C}[z_{1},z_{2}, \ldots,z_{n}]$ and the complement of a Zariski closed set is called a Zariski open set.

Given a
set of $q$ polynomials $\{g_{1},g_{2},\ldots,g_{q}\}$ with $g_{i}\in
\mathbb{C}[z]$, we define a complex affine variety as
\begin{eqnarray}
&&\mathcal{V}_{\mathbb{C}}(g_{1},g_{2},\ldots,g_{q})=\{P\in\mathbb{C}^{n}:
g_{i}(P)=0,~\text{for all}~1\leq i\leq q\}.
\end{eqnarray}
For example the space $\mathbb{C}^{n}$, the empty set and one-point
sets are trivial affine algebraic varieties given by
$\mathcal{V}_{\mathbb{C}}(0)=\mathbb{C}^{n}$,
$\mathcal{V}_{\mathbb{C}}(1)=\emptyset$, and
\begin{eqnarray}
&&\mathcal{V}_{\mathbb{C}}(z_{1}-a_{1},z_{2}-a_{2},\ldots,z_{n}-a_{n})=\{(a_{1},a_{2},\ldots,a_{n})\}.
\end{eqnarray}
Let $\mathcal{V}_{\mathbb{C}}$ be complex affine algebraic variety.
Then an ideal of $\mathbb{C}[z_{1},z_{2}, \ldots,z_{n}]$ is defined
by
\begin{eqnarray}
&&\mathcal{I}(\mathcal{V}_{\mathbb{C}})=\{g\in\mathbb{C}[z_{1},z_{2},
\ldots,z_{n}]: g(z)=0,~\text{for all}~z\in\mathcal{V}_{\mathbb{C}}\}.
\end{eqnarray}
 Note also that
 $\mathcal{V}_{\mathbb{C}}(\mathcal{I}(\mathcal{V}_{\mathbb{C}}))=\mathcal{V}_{\mathbb{C}}$.
 Moreover, we define a coordinate ring of an affine variety $\mathcal{V}_{\mathbb{C}}$
 by $C[\mathcal{V}_{\mathbb{C}}]=\mathbb{C}[z_{1},z_{2},
 \ldots,z_{n}]/\mathcal{I}(\mathcal{V}_{\mathbb{C}})$.
 A complex projective space $\mathbb{CP}^{n}$ is
defined to be the set of lines through the origin in
$\mathbb{C}^{n+1}$, that is, $
\mathbb{CP}^{n}=\frac{\mathbb{C}^{n+1}-\{0\}}{
(x_{1},\ldots,x_{n+1})\sim(y_{1},\ldots,y_{n+1})},~\lambda\in
\mathbb{C}-0,~y_{i}=\lambda x_{i}$ for all $0\leq i\leq n+1 $. For
example $\mathbb{P}_{\mathbb{C}}^{1}=\mathbb{C}\cup\{\infty\}$,
$\mathbb{P}_{\mathbb{C}}^{2}=\mathbb{C}^{2}\cup\mathbb{P}_{\mathbb{C}}^{1}=\mathbb{C}^{2}\cup\mathbb{C}\cup\{\infty\}$
and in general we have
$\mathbb{CP}^{n}=\mathbb{C}^{n}\cup\mathbb{CP}^{n-1}$.

Given a set of homogeneous polynomials
$\{g_{1},g_{2},\ldots,g_{q}\}$ with $g_{i}\in C[z]$, we define a
complex projective variety as
\begin{equation}
\mathcal{V}(g_{1},\ldots,g_{q})=\{O\in\mathbb{CP}^{n}:
g_{i}(O)=0~\forall~1\leq i\leq q\},
\end{equation}
where $O=[a_{1},a_{2},\ldots,a_{n+1}]$ denotes the equivalent class
of point $\{\alpha_{1},\alpha_{2},\ldots,$
$\alpha_{n+1}\}\in\mathbb{C}^{n+1}$. We can view the affine complex
variety
$\mathcal{V}_{\mathbb{C}}(g_{1},g_{2},\ldots,g_{q})\subset\mathbb{C}^{n+1}$
as a complex cone over the complex projective variety
$\mathcal{V}(g_{1},g_{2},\ldots,g_{q})$. The ideal and coordinate
ring of complex projective variety can be defined in similar way as
in the case of complex algebraic affine variety by considering the
complex projective space and its homogeneous coordinate.
The Zariski topology on $\mathbb{C}^{n}$ is defined to be the topology whose  closed sets are the set
\begin{equation}
\mathcal{V}(\mathcal{I})=\{z\in \mathbb{C}^{n}:g(z)=0,~\text{for all}~g\in\mathcal{I}\},
\end{equation}
where $I\subset\mathbb{C}[z_{1},z_{2},
 \ldots,z_{n}]$ is an ideal.
Let $\mathcal{A}$ be a finitely generated $\mathbb{C}$-algebra without zero divisors. Then an ideal $\mathcal{I}$ in $\mathcal{A}$ is called prime ideal if for $f,g\in\mathcal{A}$ and $f,g\in \mathcal{I}$ implies that $f\in \mathcal{I}$ or $g\in \mathcal{I}$. An ideal $\mathcal{I}\subset\mathcal{A}$ is called maximal if $\mathcal{I}\neq\mathcal{A}$ and the only proper ideal in $\mathcal{A}$ containing $\mathcal{I}$ is $\mathcal{I}$. The spectrum of the algebra $\mathcal{A}$ is the set $\mathrm{Spec} \mathcal{A}=\{\mathrm{prime}~ \mathrm{ideal}~ \mathrm{in} ~\mathcal{A}\}$ equipped with the Zariski topology. The maximal spectrum of $\mathrm{Specm} \mathcal{A}=\{\mathrm{maximal}~ \mathrm{ideal}~ \mathrm{in} ~\mathcal{A}\}$ equipped with the Zariski topology. Let us consider $\mathcal{A}=\mathbb{C}[z_{1},z_{2},
 \ldots,z_{n}]$ and let $a=(a_{1},a_{2},\ldots,a_{n})\in\mathbb{C}^{n}$. Then associated to the point $a$ there is a maximal ideal $\mathcal{I}_{a}\subset \mathcal{A}$ consisting of all polynomials which vanish at $a$, that is
 \begin{equation}\mathcal{I}_{a}=\langle z_{1}-a_{1},z_{2}-a_{2},\ldots,z_{n}-a_{n}\rangle.
\end{equation}
 Moreover any maximal ideal of $\mathbb{C}[z_{1},z_{2},
 \ldots,z_{n}]$ is of the form $\mathcal{I}_{a}$ for some $a\in\mathbb{C}^{n}$. Furthermore the correspondence $\mathbb{C}^{n}\simeq\mathrm{Specm} \mathbb{C}[z_{1},z_{2},
 \ldots,z_{n}]$ is a homeomorphism for Zariski topology. Now, let  $\mathbb{X}$ be an affine variety in $\mathbb{C}^{n}$ defined by polynomials $g_{1},g_{2},\ldots, g_{r}$ from the polynomial ring $\mathbb{C}[z_{1},z_{2},
 \ldots,z_{n}]$ and $\mathcal{I}=\langle g_{1},g_{2},\ldots, g_{r}\rangle$. Then we have $\mathbb{X}\simeq\mathrm{Specm} \mathbb{C}[\mathbb{X}]$. If $\mathcal{A}$ is finitely generated $\mathbb{C}$-algebra without zero divisors, then we called $\mathbb{X}_{\mathcal{A}}\simeq\mathrm{Specm} \mathbb{C}[\mathbb{X}]$ an abstract affine variety. Note that in this construction maximal ideals are points in $\mathbb{X}_{\mathcal{A}}$ and prime ideals are irreducible subvarieties.
Finally, we also need to define a rational mapping.
 A  map $\phi:\mathbb{X}\longrightarrow\mathbb{Y}$ is called rational if we have a  morphism from a Zariski open set $U\subset\mathbb{X}$ to $\mathbb{Y}$, that is $\phi: U\longrightarrow\mathbb{Y}$.
\subsection{Segre variety}
In this susbsection we will review the
construction of the  Segre variety \cite{Hosh1} for general multi-projective complex space.  Let
$(\alpha^{j}_{1},\alpha^{j}_{2},\ldots,\alpha^{j}_{N_{j}})$  be
points defined on the complex projective space
$\mathbb{CP}^{N_{j}-1}$. Then the Segre map
\begin{equation}
\begin{array}{ccc}
  \mathcal{S}_{N_{1},\ldots,N_{m}}:\mathbb{CP}^{N_{1}-1}\times\mathbb{CP}^{N_{2}-1}
\times\cdots\times\mathbb{CP}^{N_{m}-1}&\longrightarrow&
\mathbb{CP}^{N_{1}N_{2}\cdots N_{m}-1}
\end{array}
\end{equation}
is defined by $
 ((\alpha^{1}_{1},\alpha^{1}_{2},\ldots,\alpha^{1}_{N_{1}}),\ldots,
 (\alpha^{m}_{1},\alpha^{m}_{2},\ldots,\alpha^{m}_{N_{m}}))  \longmapsto
 (\alpha^{1}_{i_{1}}\alpha^{2}_{i_{2}}\cdots\alpha^{m}_{i_{m}})$. Now, we let $\alpha_{i_{1}i_{2}\cdots i_{m}}$,$1\leq
i_{j}\leq N_{j}$ be a homogeneous coordinate-function on
$\mathbb{CP}^{N_{1}N_{2}\cdots N_{m}-1}$. Moreover,  let $
\mathcal{A}=\left(\alpha_{i_{1}i_{2}\ldots i_{m}}\right)_{1\leq
i_{j}\leq N_{j}}, $ for all $j=1,2,\ldots,m$. $\mathcal{A}$ can be
realized as the following set $\{(i_{1},i_{2},\ldots,i_{m}):1\leq
i_{j}\leq N_{j},\forall~j\}$, in which each point
$(i_{1},i_{2},\ldots,i_{m})$ is assigned the value
$\alpha_{i_{1}i_{2}\ldots i_{m}}$. This realization of $\mathcal{A}$
is called an $m$-dimensional box-shape matrix of size $N_{1}\times
N_{2}\times\cdots\times N_{m}$, where we associate to each such
matrix a sub-ring
$\mathrm{S}_{\mathcal{A}}=\mathbb{C}[\mathcal{A}]\subset\mathrm{S}$,
where $\mathrm{S}$ is a commutative ring over the complex number
field. For each $j=1,2,\ldots,m$, a two-by-two minor about the
$j$-th coordinate of $\mathcal{A}$ is given by
\begin{equation}\label{segreply1}
\alpha_{k_{1}k_{2}\ldots k_{m}}\alpha_{l_{1}l_{2}\ldots l_{m}} -
\alpha_{k_{1}k_{2}\ldots k_{j-1}l_{j}k_{j+1}\ldots
k_{m}}\alpha_{l_{1}l_{2} \ldots l_{j-1} k_{j}l_{j+1}\ldots l_{m}}\in
\mathrm{S}_{\mathcal{A}}.
\end{equation}
Then the ideal $\mathcal{I}_{\mathcal{A}}$ of
$\mathrm{S}_{\mathcal{A}}$ is generated by this equation
 and
describes the separable states in $\mathbb{CP}^{N_{1}N_{2}\cdots
N_{m}-1}$. The image of the Segre embedding
$\mathrm{Im}(\mathcal{S}_{N_{1},N_{2},\ldots,N_{m}})$, which again
is an intersection of families of  hypersurfaces in
$\mathbb{CP}^{N_{1}N_{2}\cdots N_{m}-1}$, is called Segre variety
and it is given by
\begin{equation}\label{eq: submeasure}
\mathrm{Im}(\mathcal{S}_{N_{1},\ldots,N_{m}})=\bigcap_{\forall
j}\mathcal{V}(\alpha_{k_{1}\ldots
k_{m}}\alpha_{l_{1}\ldots l_{m}}- \alpha_{k_{1}\ldots
k_{j-1}l_{j}k_{j+1}\ldots k_{m}}\alpha_{l_{1} \ldots l_{j-1}
k_{j}l_{j+1}\ldots l_{m}}).
\end{equation}
In following sections we will construct the toric varieties and projective toric varieties of the Segre varieties with application to the geometry of quantum systems.
\section{Toric variety}\label{toric}
This section  is very technical. First we recall some basic definitions and facts from combinatorial geometry and then  we will define toric varieties. Finally we show the construction of projective toric varieties by embedding it to a complex projective space. The main reference for this section is the book by G. Ewald \cite{Ewald}.

\subsection{Combinatorial geometry}
Convexity and convex functions play an important role in physics and engineering. Here we are mostly interested in combinatorial structure convexity. The most important combinatorial structures for construction of toric varieties are polytope, fan, and monoid which we will define in following text.
 We call $x$ a convex combination of $x_{1},x_{2},\ldots,x_{k}\in \mathbb{R}^{n}$ if there exist $\lambda_{1},\lambda_{2}, \ldots, \lambda_{k}\in \mathbb{R}$ such that $x=\lambda_{1}x_{1}+\lambda_{2}x_{2}+\cdots+\lambda_{k}x_{k}$, with $\sum^{r}_{i=1}\lambda_{i}=1$ for all $\lambda_{i}\geq0$ and $i=1,2,\ldots, k$. If the convex combination of $M$ does not satisfies the condition $\lambda_{1}\geq0,\lambda_{2}\geq0, \ldots, \lambda_{k}\geq0$ then we call $x_{1},x_{2},\ldots,x_{k}$ an affine combination. The set of all convex combination of elements of the set $M\in \mathbb{R}^{n}$ is called the convex hull $\mathrm{conv}M$ of $M$ and the set of all affine combinations of elements of $M$ is called the affine hull $\mathrm{aff}M$ of $M$. The linear space generated by $M$ is also denoted by $\mathrm{lin}$. Moreover, if $M$ is a finite set, then we call $\mathrm{conv}M$ a convex polytope. Note that $M\subset\mathrm{conv}M\subset \mathrm{aff}M$ and every polytope is  compact. We also say $x\in M\in \mathbb{R}^{n}$ is in the relative interior of $M$, $x\in\mathrm{relint}M$ if $x$ is the interior of the $M$ relative to $\mathrm{aff} M$.

Let $M\subset \mathbb{R}^{n}$. Then the set of all nonnegative linear combinations $x=\lambda_{1}y_{1}+\lambda_{2}y_{2}+\cdots+\lambda_{k}y_{k}$, where $y_{1},y_{2},\ldots,y_{k}\in M$ and $\lambda_{i}\geq0$ for $i=1,2,\ldots, k$ of elements of $M$ is called the positive hull $\Delta=\mathrm{pos}M$ of $M$ or the cone determined by $M$. If $M=\{x_{1},x_{2},\ldots,x_{k}\}$ is finite, then we call
 \begin{equation}
 \Delta=\mathrm{pos}\{x_{1},x_{2},\ldots,x_{k}\}
\end{equation}
a polyhedral cone. Moreover, we call $\Delta=\mathrm{pos}\{x_{1},x_{2},\ldots,x_{k}\}$ a simplex cone if $(x_{1},x_{2},\ldots,x_{k})$  are linearly independent and it is said to be simplicial if each proper face of $\Delta$ is a simplex cone. We can also write this polyhedral cone as
\begin{equation}
\Delta=\mathbb{R}_{\geq0}x_{1}+\mathbb{R}_{\geq0}x_{2}+\cdots+\mathbb{R}_{\geq0}x_{r},
\end{equation}
 where $\mathbb{R}_{\geq0}$ is denoting the set of nonnegative real numbers. Note that the positive hull $\Delta=\mathrm{pos}M$ is convex. Furthermore, the cone $\Delta$ is called rational if it admits a set of generators in $\mathbb{Z}^{n}$. If 0 is a face of $\Delta$, then $\Delta$ is called strongly convex. In this case $\Delta$ contains no one-dimensional $\mathbb{R}$-subspace, that is $\Delta\cap(-\Delta)=\{0\}$. If $\Delta$ is strongly convex, then its dual is $n$-dimensional, i.e.,
 \begin{equation}\Delta+(-\Delta)=(\mathbb{R}^{n})^{*}.
 \end{equation}
 A hyperplane $H$ is called a supporting hyperplane of a closed convex set $K\subset\mathbb{R}^{n}$ if $K\cap H\neq0$ and $K\cap H^{+}$ or $K\cap H^{-}$, where $H^{+}$ and  $H^{-}$ are supporting half-space of $K$. If $H$ is a supporting hyperplane of closed convex set $K$, we call $F=K\cap H$ a face of $K$. Note that the $\varnothing$ the empty set and $H$ are improper faces of $K$.  Let $\Delta$ be a cone. Then the set
 \begin{equation}\check{\Delta}=\{y:(\Delta,y)\geq0\}
 \end{equation}
 is called the dual of $\Delta$. Moreover, let $M\simeq\mathbb{Z}^{n}$ and  $N=\mathrm{Hom}(M,\mathbb{Z})$ be its dual and let $N_{\mathbb{R}}\subset N\otimes \mathbb{R}$. Then the polar of $\Delta$ is defined by
  \begin{equation}\Delta^{\circ}=\{y\in N_{\mathbb{R}}:(x,y)\geq-1,~ \text{for} ~\text{all}\in\Delta\}.
 \end{equation}
 A semigroup is a non-empty set with an associative operation $+:S\times  S\longrightarrow S$. A semigroup is called  a  monoid if it is commutative and has a zero, e.g., an element $0\in S$ which satisfies $s+0=s$ for all $s\in S$ and it satisfies the cancelation law $s+x=t+x$ implies $t=s$, for all $s,t,x \in S$. Now, if $\Delta$ is a cone in $\mathbb{R}^{n}$, then $\Delta \cap \mathbb{Z}^{n}$ is a monoid. A monoid $S$ is said to be finitely generated if there exist $a_{1},a_{2},\ldots, a_{r}\in S$ such that
 \begin{equation}
 S=\mathbb{Z}_{\geq0}a_{1}+\mathbb{Z}_{\geq0}a_{2}+\cdots+\mathbb{Z}_{\geq0}a_{r},
 \end{equation}
 where  $a_{1},a_{2},\ldots, a_{r}$ are also called the generators of $S$ and a system of  these generators is called minimal if none of its elements is generated by the others.  We call the points $a\in \mathbb{Z}^{n}\subset \mathbb{R}^{n}$ a lattice vectors and if the vector $a_{1},a_{2} ,\ldots,a_{k}$ in the polyhedral cone $\Delta=\mathbb{R}_{\geq0}a_{1}+\mathbb{R}_{\geq0}a_{2}+\cdots+\mathbb{R}_{\geq0}a_{r}$ are lattice vectors, then we call $\Delta$ a lattice cone. The cone $\Delta$ is called rational if it admits a set of generators in $\mathbb{Z}^{n}$. If $\Delta$ is a lattice cone in $\mathbb{R}^{n}$, then the monoid $\Delta\cap\mathbb{Z}^{n}$ is finitely generated. A monoid  $S$ can be embedded as a subsemigroup into a group $G(S)$ which has also $a_{1},a_{2},\ldots, a_{r}$ as generators  with coefficients in $\mathbb{Z}$ .

 Now, let $K$ be a closed convex set in $\mathbb{R}^{n}$. Then to each $x\in\mathbb{R}^{n}$ there exists a unique $x'\in K$ such that
 \begin{equation}\|x-x'\|=\inf_{y\in K}\|x-y\| .
\end{equation}
 Moreover, the map $p_{K}:\mathbb{R}^{n}\longrightarrow K$ defined by $x\longmapsto p_{K}(x)=x'$ is called the nearest point map relative to $K$. The map $p_{K}$ is surjective and $p_{K}(x)=x$ if and only if $x\in K$. Next, let $x$ be a point of the closed convex set $K$. Then we call $N(x)=-x+p^{-1}_{K}(x)$ the normal cone of $K$ at $x$ which is a closed convex cone consisting of 0 and all outer normals of $K$ in $x$. If $\Delta$ is a cone with apex 0, then  $N(0)=-\check{\Delta}$. Let $F$ be a face of a closed convex set $K$ and  $x\in\mathrm{relint}F$. Then $N(x)$ is called the cone of normals of $K$ and is denoted by $N(F)$. The fan of a convex set $K$ is defined to be the set of all cones $N(F)$ and is denoted by $\Sigma(K)$. If $K$ is a polytope and has only finitely many faces. Then  $\Sigma(K)$ consists of finitely many cones. Moreover, the fan € $\Sigma=\Sigma(P)$  is said to be strongly polytopal if there exists a polytope $P^{*}$ such that $0\in P^{*}$ and  $\Sigma=\{\mathrm{pos} F: F\in\mathcal{B}(P^{*})\}$, where $\mathcal{B}(P)$ denotes the boundary of polytope $P$. $P^{*}$ is called the spanning polytope of the fan $\Sigma$. We also call $P$ an associated polytope of $\Sigma$ if $\Sigma=\Sigma(-P)$, that is if $\Sigma$ is spanned by $-P^{*}$ of $-P$.
  Given any fan $\Sigma$ we  defined its dual cone as $\check{\Sigma}=\{\check{\Delta}:\Delta\in\Sigma\}$ and if $\Delta\in\Sigma$, then we assign to $\check{\Delta}$ the monoid $S=\check{\Delta}\cap \mathbb{Z}^{n}$.
  We also have
  \begin{equation}
  S=S(\Sigma)=\{S=\check{\Delta}\cap \mathbb{Z}^{n}:\Delta\in\Sigma\}
  \end{equation}
  of monoid assigned to the fan $\Sigma$ and there are bijective relations between $\Sigma,\check{\Sigma}$, and $S(\Sigma)$, that is $\Sigma\longleftrightarrow\check{\Sigma}\longleftrightarrow S(\Sigma)$.
 Now, let $\Sigma=\Sigma(-P)$ be the fan of a polytope $-P$. Then, $\bar{P}=nP$ has the following property. For any lattice point $a$ of $\bar{P}$, the generators of the monoid $\check{\Delta}_{a}\cap\mathbb{Z}^{n}=(\mathrm{pos}(\bar{P}-a))\cap\mathbb{Z}^{n}$ lie in $\bar{P}-a$.

\subsection{Abstract toric variety}

Affine toric varieties and algebraic toric varieties have very rich mathematical structures which also have found application in algebraic construction mirror symmetry in string theory. Here we will give a short introduction to basic definition and properties of these interesting varieties.
Let $R=\mathbb{C}[\xi_{1},\xi_{2},\ldots,\xi_{2n}]$ be a polynomial ring in $2n$ variables for $n\geq 0$. Then we define an ideal $\mathcal{I}$ of $R$ as
\begin{equation}
\mathcal{I}=R(\xi_{1}\xi_{n+1}-1)+\cdots+R(\xi_{n}\xi_{2n}-1).
\end{equation}
If we set $z_{i}=\xi_{i}+\mathcal{I}\in R/\mathcal{I}$ for $i=1,2\ldots, 2n$, then we have $z_{l}z_{l+1}=1$ and $z^{-1}_{l}=z_{l+1}$ for all $l=1,2,\ldots,n$. A Laurent polynomials are defined to be the elements of
\begin{equation}
\mathbb{C}[z,z^{-1}]=\mathbb{C}[z_{1},z_{2},\ldots,z_{n},z^{-1}_{1},z^{-1}_{2},\ldots,z^{-1}_{n}]
=\mathbb{C}[\xi_{1},\xi_{2},\ldots,\xi_{2n}]/\mathcal{I}
\end{equation}
and the terms  of the form
 \begin{equation}\label{mono}
 \lambda \cdot z^{\beta}=\lambda z^{\beta_{1}}_{1}z^{\beta_{2}}_{2}\cdots z^{\beta_{n}}_{n}, ~ \text{for}~ \beta=(\beta_{1},\beta_{2},\ldots,\beta_{n})\in \mathbb{Z}
\end{equation}
and $\lambda\in \mathbb{C}^{\times}$ are called Laurent monomials. A ring $R$ of Laurent polynomials is called a monomial algebra if it is a $\mathbb{C}$-algebra generated bye Laurent monomials. Moreover, for a lattice cone $\Delta$, the ring
 \begin{equation}R_{\Delta}=\{f\in \mathbb{C}[z,z^{-1}]:\mathrm{supp}(f)\subset \Delta\}
\end{equation}
is a finitely generated monomial algebra, where the support of a Laurent polynomial $f=\sum\lambda_{i}z^{i}$ is defined by
$\mathrm{supp}(f)=\{i\in \mathbb{Z}^{n}:\lambda_{i}\neq0\}.
$
Now, for a lattice cone $\Delta$ we can define an affine toric variety to be the maximal spectrum $\mathbb{X}_{\Delta}=\mathrm{Specm}R_{\Delta}$. We can also realize $\mathbb{X}_{\Delta}$ as the set of points
\begin{equation}
T=\{(z_{1},z_{1},\ldots,z_{n})\in\mathbb{C}^{n}:z_{i}\neq0,~i=1,2,\ldots,n\}
=(\mathbb{C}^{\times})^{n}=(\mathbb{C}\backslash\{0\})^{n},
\end{equation}
Under the projection $\mathbb{C}^{2n} \longrightarrow \mathbb{C}^{n}$ this realization is also isomorphic to
$V(\xi_{1}\xi_{n+1}-1),\ldots,\xi_{n}\xi_{2n}-1)$. Moreover the inverse of the restricted projection is given by
\begin{equation}
\begin{array}{ccc}
  V(z_{1}z_{n+1}-1,\ldots,z_{n}z_{2n}-1)&\longrightarrow& T \\
  (z_{1},z_{2},\ldots,z_{n})&\longmapsto& (z_{1},z^{-1}_{1},z_{2},z^{-1}_{2}\ldots,z_{n},z^{-1}_{n}).
\end{array}
\end{equation}
The set $T$ is called a complex algebraic $n$-torus. Note that an algebraic torus is not a compact complex torus. Let $A=(a_{1},a_{2},\ldots, a_{k})$ be a system of generators of the monoid $\Delta\cap\mathbb{Z}^{n}$. If we set $u_{i}=z^{a_{i}}\in \mathbb{C}[z,z^{-1}]$, for all $i=1,2,\ldots, k$, then $R_{\Delta}=\mathbb{C}[u_{1},u_{2},\ldots, u_{k}]$. There is also an algebra homomorphism
\begin{equation}
\phi_{A}:\mathbb{C}[\xi_{1},\xi_{2},\ldots, \xi_{k}]\longrightarrow\mathbb{C}[u_{1},u_{2},\ldots, u_{k}],
\end{equation}
where $\xi_{i}\longmapsto u_{i}$ for all $i=1,2,\ldots, k$. Now, let $\mathcal{I}=\mathcal{I}_{A}=\mathrm{Ker}\phi_{A}$ be defined as follows. For every $f\in\mathbb{C}[\xi]$, for $f\in \mathcal{I}_{A}$ if and only if $\phi_{A}(f)=0$ which is equivalent to $f(u)=0$ in $\mathbb{C}[u]$. Moreover, we have
\begin{equation}
R_{\Delta}=\mathbb{C}[u_{1},u_{2},\ldots, u_{k}]\cong\mathbb{C}[\xi_{1},\xi_{2},\ldots, \xi_{k}]/\mathcal{I}_{A}.
\end{equation}
Furthermore, for every lattice cone $\Delta$, the toric variety $\mathbb{X}_{\Delta}$ is realized as by the affine algebraic variety $\mathcal{V}(\mathcal{I}_{A})$ in $\mathbb{C}^{n}$ and the ideal $\mathcal{I}_{A}$ of $\mathbb{C}[\xi_{1},\xi_{2},\ldots, \xi_{k}]$ is generated by finitely many binomials of the form $\xi^{\nu}-\xi^{\mu}$ which is constructed as follows. For $\mu,\nu\in \mathbb{Z}^{k}_{\geq0}$ and the system of generators $A$ we have $\mu\cdot A=\nu\cdot A$, where e.g., $\nu\cdot A=\sum^{k}_{l=1}\nu_{l}A_{l}$. This relations give a monomial equation
\begin{equation}
z^{\nu\cdot A}=(z^{a_{1}})^{\nu_{1}}(z^{a_{2}})^{\nu_{2}}\cdots(z^{a_{k}})^{\nu_{k}}=
(z^{a_{1}})^{\mu_{1}}(z^{a_{2}})^{\mu_{2}}\cdots(z^{a_{k}})^{\mu_{k}}=z^{\mu\cdot A}.
\end{equation}
Moreover, we can write this relation in terms of generators $u_{l}=z^{a_{l}}$ of $R_{\Delta}$ as a binomial relationship
\begin{equation}
 u^{\nu_{1}}_{1}u^{\nu_{2}}_{2}\cdots u^{\nu_{k}}_{k}=
u^{\mu_{1}}_{1}u^{\mu_{2}}_{2}\cdots u^{\mu_{k}}_{k}.
\end{equation}
Thus for a relationship $(\nu,\mu)$ the corresponding  binomials are in the form $\xi^{\nu}-\xi^{\mu}$.
For lattice cone $\Delta\subset \mathbb{R}^{n_{1}}=\mathrm{lin}\Delta$ and $\Delta'\subset \mathbb{R}^{n_{2}}=\mathrm{lin}\Delta'$, the following conditions are equivalent:
\begin{equation}
 \Delta=\Delta'\Leftrightarrow R_{\Delta}=R_{\Delta'}\Leftrightarrow \mathbb{X}_{\Delta}=\mathbb{X}_{\Delta'}.
\end{equation}
Now, we can construct general toric variety by gluing affine toric varieties. If $\tau$ is a face of a cone $\Delta$, then we have an inclusion $\check{\Delta}\subset \check{\tau}$ and the induced toric morphism $\psi_{\tau,\Delta}:\mathbb{X}_{\check{\tau}}\longrightarrow\mathbb{X}_{\check{\Delta}}$ is an isomorphism given by a coordinate transformation on the copies of the complex algebraic $n$-torus $T$. If $\psi_{\tau,\Delta}$ is an open inclusion, then gluing any two such affine toric varieties $\mathbb{X}_{\check{\Delta}},\mathbb{X}_{\check{\Delta}'}$ along the open subsets corresponding to the common face $\tau=\Delta\cap\Delta'$. Now, let us choose a lattice vector $l\in \mathrm{relint}(\tau^{\perp}\cap\check{\Delta})$ which satisfies $\check{\tau}=\check{\Delta}+\mathbb{R}_{\geq0}(-l)$ and
\begin{equation}\check{\tau}\cap\mathbb{Z}^{n}=(\Delta\cap\mathbb{Z}^{n})+\mathbb{Z}_{\geq0}(-l).
\end{equation}
Let $a_{1},a_{2},\ldots,a_{k}$ be generator of $\check{\Delta}\cap\mathbb{Z}^{n}$ and $u_{i}=z^{a_{i}}$ be the coordinate functions on the toric varieties $\mathbb{X}_{\check{\Delta}}$ and $\mathbb{X}_{\check{\tau}}$. Then we have $\mathbb{X}_{\check{\tau}}\cong\mathbb{X}_{\check{\Delta}}\backslash\{u_{k}=0\}$. For two cones $\Delta,\Delta'\in\Sigma$, let
 $\tau=\Delta\cap\Delta'$ be the common face. If $v_{1},v_{2},\ldots,v_{q}$ are the coordinate system for $\mathbb{X}_{\check{\Delta'}}$, then we have following isomorphisms
 \begin{equation}
 \mathbb{X}_{\check{\Delta}}\backslash\{u_{k}=0\}\cong\mathbb{X}_{\check{\tau}}\cong\mathbb{X}_{\check{\Delta'}}\backslash\{v_{k}=0\}.
 \end{equation}
 Moreover, we have an isomorphism
\begin{equation}
 \begin{array}{ccc}
   \psi_{\Delta,\Delta'}:\mathbb{X}_{\check{\Delta}}\backslash\{u_{k}=0\}& \longrightarrow & \mathbb{X}_{\check{\Delta'}}\backslash\{v_{k}=0\} \\
   (u_{1},u_{2},\ldots,u_{k},u_{k+1}) & \longmapsto & (v_{1},v_{2},\ldots,v_{q},v_{q+1}).
 \end{array}
\end{equation}
The isomorphism map $\psi_{\Delta,\Delta'}$ is called the gluing map that glues together $\mathbb{X}_{\check{\Delta}}$ and $\mathbb{X}_{\check{\Delta'}}$ along $\mathbb{X}_{\check{\tau}}$. After all these technicality we are now in position to define a toric variety as the set of following points. Let $\Sigma$ be a fan in $\mathbb{R}^{n}$. Then in the disjoint union $\bigcup_{\Delta\in \Sigma}\mathbb{X}_{\check{\Delta}}$ we identify two points $x\in\mathbb{X}_{\check{\Delta}}$ and $x'\in\mathbb{X}_{\check{\Delta'}}$ that mapped to each other by the map $\psi_{\Delta,\Delta'}$. The set of point that is obtained in this way is called the toric variety $\mathbb{X}_{\Sigma}$ determined by the fan $\Sigma$.

Let $\Delta$ be a cone with apex $0\in\mathbb{R}$ and $\tau$ be a proper face of the dual cone $\check{\Delta}$. Moreover, let $a_{1},a_{2},\ldots a_{k}$ be a system of generators of the monoid $\tau\cap \mathbb{Z}^{n}$. Then we can extend this system of generators to $a_{1},a_{2},\ldots a_{k},a_{k+1},a_{k+2},\ldots, a_{q}$ of the generators of $\check{\Delta}\cap \mathbb{Z}^{n}$. Such system provides coordinates $(u_{1},u_{2},\ldots u_{k})$ and $(u_{1},u_{2},\ldots u_{k},u_{k+1},u_{k+2},\ldots, u_{q})$ with $u_{i}=z^{a_{i}}$ for $i=1,2,\ldots, q$ for affine varieties $\mathbb{X}_{\tau}$ and $\mathbb{X}_{\check{\Delta}}$. If we assume that $a_{k+1},a_{k+2},\ldots, a_{q}$ are not belong to $\tau\cap \mathbb{Z}^{n}$, then none of the coordinate functions $u_{k+1},u_{k+2},\ldots, u_{q}$ is invertible on $\mathbb{X}_{\check{\Delta}}$ and there is also a natural mapping
$\phi:\mathbb{X}_{\check{\tau}}\longrightarrow\mathbb{X}_{\check{\Delta}}$ defined by $(u_{1},u_{2},\ldots u_{k})\longmapsto(u_{1},u_{2},\ldots u_{k},0,0,\ldots,0)$. This map is an injective affine toric morphism. Thus we can identify $\mathbb{X}_{\tau}$ and the closed subvariety $\phi(\mathbb{X}_{\tau})=\mathbb{X}_{\check{\Delta}}\cap(u_{k+1}=u_{k+2}=\cdots= u_{q}=0)$. For example let $\Delta=\check{\Delta}=\mathrm{pos}\{e_{1},e_{2}\}$ be the first quadrant in $\mathbb{R}^{2}$ and $\tau=\mathrm{pos}\{e_{1}\}$. Then $\phi(\tau)$ is the $\xi_{1}$ axis of the affine plane $\mathbb{X}_{\check{\Delta}}=\mathbb{C}^{2}$.

\subsection{Projective toric variety}
Projective variety is defined by embedding an algebraic toric variety into a complex projective variety. The projective toric  varieties of the Segre varieties give us  very simple way of constructing the  space of separable states of multipartite states.
In this subsection we give short introduction to the embedding of a toric variety into a projective space $\mathbb{CP}^{r}$.

If  any orbit is the embedding $T_{k}$ of a torus of some dimension$k$ for $0\leq k\leq n$. Then we call each $T_{k}$ an embedding torus and $T_{k}=T$ is said to be the big torus in a toric variety $\mathbb{X}_{\Sigma}$.
A compact toric variety $\mathbb{X}_{\Sigma}$ is called projective if there exists an injective morphism
$\Phi:\mathbb{X}_{\Sigma}\longrightarrow\mathbb{CP}^{r}$ of $\mathbb{X}_{\Sigma}$ into some projective space such that $\Phi(\mathbb{X}_{\Sigma})$ is Zariski closed in $\mathbb{CP}^{r}$. We can also consider $\mathbb{CP}^{r}$ as a toric variety whose big torus $T$ has dimension $r$. Moreover, $\Phi$ is said to be equivariant if it is equivariant in following sense. Let $\mathbb{X}_{\Sigma}$ and $\mathbb{X}_{\Sigma'}$ be toric varieties with embedded tori $T\subset\mathbb{X}_{\Sigma}$ and $T'\subset\mathbb{X}_{\Sigma'}$ and also let $\Theta:\mathbb{X}_{\Sigma}\longrightarrow\mathbb{X}_{\Sigma'}$ be a map and $\vartheta:T\longrightarrow T'$ be a homomorphism such that $\Theta(w\cdot x)=\vartheta(w)\cdot \Theta(x)$. Then we call $\Theta$ equivariant with respect to $\vartheta$. In this case the image $\Phi(T_{0})$ of the big torus $T_{0}$ acts on $\Phi(\mathbb{X}_{\Sigma})$  as a subgroup of $T$. $\mathbb{X}_{\Sigma}$  is called to be equivariantly projective if the mapping $\Phi$ is equivariant. Now, let $\phi:\mathbb{X}_{\Sigma}\longrightarrow\mathbb{CP}^{r}$ be an equivariant morphism. Then we have
\begin{equation}
\phi|_{T_{0}}:T_{0}\longrightarrow T=\{[x_{0},\ldots, x_{r}]:x_{i}\neq0, i=0,1,\ldots,r\}
\end{equation}
defined by $z=(z_{1},z_{2},\ldots,z_{n})\longrightarrow[z^{m_{0}},z^{m_{1}},\ldots,z^{m_{r}}]$ and the monomial $z^{m_{i}}$ are defined up to a common multiple $z^{m}$. Thus we can choose $z^{m_{0}+m},z^{m_{1}+m},\ldots,$ $z^{m_{r}+m}$ as a representative. A rational map defined by $(z^{m_{0}},z^{m_{1}},\ldots,z^{m_{r}})$ is a morphism if and only if for each $p\in \mathbb{X}_{\Sigma}$ we  can find a monomial $z^{m}$ such that the monomials $z^{m_{0}+m},z^{m_{1}+m},\ldots,z^{m_{r}+m}$ are regular at $p$ and do not vanish there.

A toric variety $\mathbb{X}_{\Sigma}$ is equivariantly projective if and only if $\Sigma$ is strongly polytopal. Now, let $\mathbb{X}_{\Sigma}$ be equivariantly projective and morphism
$\Phi:\mathbb{X}_{\Sigma}\longrightarrow\mathbb{CP}^{r}$  be embedding which is induced by the rational map $\phi$
\begin{equation}
\begin{array}{ccc}
  \phi:\mathbb{X}_{\Sigma} & \longrightarrow & \mathbb{CP}^{r} \\
  p & \longmapsto & [z^{m_{0}},z^{m_{1}},\ldots,z^{m_{r}}],
\end{array}
\end{equation}
where $z^{m_{l}}(p)=p^{m_{l}}$ in case $p=(p_{1},p_{2},\ldots p_{n})\in T$. Then the rational map $\Phi(\mathbb{X}_{\Sigma})$ is the set of common solutions of finitely many monomial equations
\begin{equation}
x^{\beta_{0}}_{i_{0}}x^{\beta_{1}}_{i_{1}}\cdots x^{\beta_{k}}_{i_{k}}=x^{\beta_{k+1}}_{i_{k+1}}x^{\beta_{k+2}}_{i_{k+2}}\cdots x^{\beta_{r}}_{i_{r}}
\end{equation}
which arise from affine relationships
\begin{equation}
\begin{array}{c}
  \beta_{0}m_{0}+\beta_{1}m_{1}+\cdots +\beta_{k}m_{k}=\beta_{k+1}m_{k+1}+\beta_{k+2}m_{k+2}+\cdots +\beta_{r}m_{r} \\
  \beta_{0}+\beta_{1}+\cdots +\beta_{k}=\beta_{k+1}+\beta_{k+2}+\cdots +\beta_{r},
\end{array}
\end{equation}
for all $\beta_{l}\in \mathbb{Z}_{\geq 0}$ and $l=0,1,\ldots, r$. After these introductory sections, we are going to apply these advanced tools to geometrical construction  of multipartite quantum systems.
\section{Toric variety and quantum system}\label{tvqs}
 In this section we will introduce a very simple way of representing complex Hilbert space and composite structure of bipartite and multipartite quantum system based on algebraic toric varieties and projective toric varieties. First, we discuss the construction of algebraic toric variety which corresponds to a complex projective  Hilbert space of a pure state and then we in detail  describe the toric variety construction of a single qubit. Moreover, we construct algebraic toric variety of two qubit and three-qubit states. Finally we also generalize this construction to multi-qubit state.
Let us denote a general, multipartite quantum system with $m$
subsystems by
$\mathcal{Q}=\mathcal{Q}_{m}(N_{1},N_{2},\ldots,N_{m})$ $
=\mathcal{Q}_{1}\mathcal{Q}_{2}\cdots\mathcal{Q}_{m}$, consisting
of a state
\begin{equation}\label{Mstate}
\ket{\Psi}=\sum^{N_{1}-1}_{k_{1}=0}\sum^{N_{2}-1}_{k_{2}=0}\cdots\sum^{N_{m}-1}_{k_{m}=0}
\alpha_{k_{1}k_{2}\cdots k_{m}} \ket{k_{1}k_{2}\cdots k_{m}}
\end{equation}
 and, let
$\rho_{\mathcal{Q}}=\sum^{\mathrm{N}}_{i=1}p_{i}\ket{\Psi_{i}}\bra{\Psi_{i}}$,
for all $0\leq p_{i}\leq 1$ and $\sum^{\mathrm{N}}_{i=1}p_{i}=1$,
denote a density operator acting on the Hilbert space $
\mathcal{H}_{\mathcal{Q}}=\mathcal{H}_{\mathcal{Q}_{1}}\otimes
\mathcal{H}_{\mathcal{Q}_{2}}\otimes\cdots\otimes\mathcal{H}_{\mathcal{Q}_{m}},
$ where the dimension of the $j$th Hilbert space is given  by
$\dim(\mathcal{H}_{\mathcal{Q}_{j}})=N_{j}-1$.

First, we will discuss the most important space in quantum physics namely the complex projective space as the space of a pure state $\ket{\Psi_{j}}=\sum^{N_{j}-1}_{k_{j}=0}
\alpha_{k_{j}} \ket{k_{j}}$ for all $j=1,2,\ldots, m$. Let
\begin{equation}
U_{i}=\{(\alpha_{1},\alpha_{2},\ldots,\alpha_{N_{j}-1})\in\mathbb{P}^{N_{j}-2}:\alpha_{k_{i}}\neq0\}
\end{equation}
 be a subset of $\mathbb{P}^{N_{j}-2}$ which can be identified with the affine complex space $\mathbb{C}^{N_{j}-2}$ by the following bijective map
\begin{equation}
\begin{array}{ccc}
  U_{i}&\longrightarrow & \mathbb{C}^{N_{j}-2}\\
  (\alpha_{1},\ldots,\alpha_{N_{j}-1}) & \longmapsto &(\alpha_{1}/\alpha_{i},\ldots,\alpha_{(i-1)}/\alpha_{i},\alpha_{(i+1)}
  /\alpha_{i},\ldots,\alpha_{N_{j}-1}/\alpha_{i})
\end{array}
\end{equation}
that defines the $i$th system of inhomogeneous coordinates  $(\zeta_{i,k})_{k=1,2,\ldots,N_{j}-2}$ on $\mathbb{P}^{N_{j}-2}$
 endowed with a covering by $N_{j}-1$ copies of the $\mathbb{C}^{N_{j}-2}$. Moreover, for $\leq i\leq l\leq N_{j}-2$, the transition from he coordinates  $(\zeta_{i,k})$ to $(\zeta_{l,k})$ that provides the gluing of $U_{i}$ and $U_{l}$ is give by a monomial transformation. The intersection $\bigcap^{N_{j}-1}_{k=1}U_{k}$ is identified with $(\mathbb{C}^{\times})^{N_{j}-2}$ and the torus $T$ is embedded in $\mathbb{P}^{N_{j}-2}$. The torus action on $\mathbb{C}^{N_{j}-2}=U_{0}$ by componentwise multiplication extends to $\mathbb{P}^{N_{j}-2}$ by
 \begin{equation}
 ((t_{1},t_{2},\ldots,t_{N_{j}-2}),(\alpha_{1},\alpha_{2},\ldots,\alpha_{N_{j}-1}))
 \longmapsto(\alpha_{1},t_{1}\alpha_{2},\ldots,t_{N_{j}-2}\alpha_{N_{j}-1}).
 \end{equation}
 Moreover, $\mathbb{P}^{N_{j}-2}$ is a toric variety determined by the fan $\Sigma$ with the $N_{j}-2$-dimensional cones $\Delta_{0}=\mathrm{pos}\{e_{1},e_{2},\ldots,e_{N_{j}-2}\}$ and
 \begin{equation}
 \Delta_{i}=\mathrm{pos}\{e_{1},\ldots,e_{i-1},e_{i+1},\ldots,e_{N_{j}-2},-(e_{1}+e_{2}
 +\cdots+e_{N_{j}-2})\}.
 \end{equation}
 The $N_{j}-1$ toric variety $\mathbb{X}_{\check{\Delta}_{i}}$ for $i=1,2,\ldots,N_{j}-2$, covering $\mathbb{X}_{\Sigma}$, are the copies $U_{i}\cong\mathbb{C}^{N_{j}-2}$ of the affine $N_{j}-2$-space that correspond to system of inhomogeneous coordinates. If $\Sigma$  has generators $e_{1},e_{2}\ldots,e_{N_{j}-2},-e_{1}-e_{2}-\cdots-e_{N_{j}-2}$, then we choose an associated polytope, the simplex
 \begin{equation}P=\mathrm{conv}\{0,-e_{1},-e_{2},\cdots,-e_{N_{j}-2}\}.
\end{equation}
The vertices of $P$ are the only lattice points of $P$ and are no affine relations between them. Thus each point $(\alpha_{1},\alpha_{2},\ldots,\alpha_{N_{j}-1})$ of $\mathbb{P}^{N_{j}-2}$ represents a point of $\mathbb{X}_{\Sigma}$ and we have $\mathbb{X}_{\Sigma}\cong\mathbb{P}^{N_{j}-2}$.
In general the toric variety associated to the standard $N$-simplex in $\mathbb{R}^{N}$ is $\mathbb{CP}^{N}$.

In following text we will discuss  the composite multipartite system as product of complex projective space $\mathbb{P}^{N_{j}-2}$.
\begin{figure}
  \includegraphics[width=12cm,height=4cm,keepaspectratio,clip]{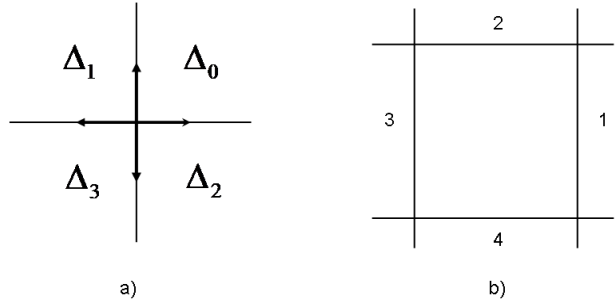}\\
  \caption{The fan of toric variety $\mathbb{X}_{\Sigma}=\mathbb{CP}^{1}\times\mathbb{CP}^{1}$ and its invariant subvarieties.}\label{Figure1}
\end{figure}
For example for a single qubit state  $\ket{\Psi}=
\alpha_{1} \ket{1}+\alpha_{2} \ket{2}$ we have following construction.
 If $\Delta\subset\mathbb{R}$, then $\mathbb{C}[\alpha_{1},\alpha_{2}]$ is a polynomial ring in two variable of degree one and the associated toric variety is the complex projective space $\mathbb{CP}^{1}$. Note that the $\mathbb{C}^{\times}$ acts on $\mathbb{C}^{2}=\mathbb{CP}^{1}$ by scalar. To see this let $z=\alpha_{2}\alpha^{-1}_{2}$
 consider the fan $\Sigma$ consisting of the three cones $\Delta=\{0\}$, $\Delta_{0}=\mathrm{pos}\{e_{1}\}$, and $\Delta_{1}=\mathrm{pos}\{-e_{1}\}$. In this case each 1-dimensional cone represents the variety $\mathbb{C}$, that is $\mathbb{C}[S_{\Delta_{0}}]=\mathbb{C}[z]$ and $\mathbb{C}[S_{\Delta_{1}}]=\mathbb{C}[z^{-1}]$. The gluing of these 1-dimensional charts is prescribes by zero dimensional cone $\Delta$ representing $\mathbb{C}^{\times}$, that is $\mathbb{C}[S_{\Delta}]=\mathbb{C}[z,z^{-1}]$. In $\mathbb{X}_{\Delta_{0}}$, the subset $\mathbb{X}_{\Delta}$ correspond to $\mathbb{C}^{*}_{z}=\{z\in\mathbb{C}:z\neq0\}$, and in $\mathbb{X}_{\Delta_{1}}$, the subset $\mathbb{X}_{\Delta}$ correspond to $\mathbb{C}^{*}_{z^{-1}}=\{z^{-1}\in\mathbb{C}:z^{-1}\neq0\}$. Now, these three cones form a fan and the corresponding toric variety is constructed from the gluing
\begin{equation}
\begin{array}{ccc}
 \mathbb{C}[z^{-1}]\hookrightarrow\mathbb{C}[z,z^{-1}]\hookleftarrow\mathbb{C}[z] \\
  \mathbb{C}\hookleftarrow\mathbb{C}^{\times}\hookrightarrow\mathbb{C}
\end{array}
\end{equation}
by using the gluing map $z\longmapsto z^{-1}$.
Thus we have constructed the toric variety $\mathbb{X}_{\Sigma}=\mathbb{CP}^{1}$ for a  single qubit state.

For a pairs of qubits  $\ket{\Psi}=\sum^{1}_{k_{1}=0}\sum^{1}_{k_{2}=0}
\alpha_{k_{1}k_{2}} \ket{k_{1}k_{2}}$ we can also construct following simplex. For this two qubit state the separable state is given by the Segre embedding of $\mathbb{CP}^{1}\times\mathbb{CP}^{1}=
\{((\alpha^{1}_{0},\alpha^{1}_{1}),(\alpha^{2}_{0},\alpha^{2}_{1})): (\alpha^{1}_{0},\alpha^{1}_{1})\neq0,~(\alpha^{2}_{0},\alpha^{2}_{1})\neq0\}$. Let $z_{1}=\alpha^{1}_{1}/\alpha^{1}_{0}$ and $z_{2}=\alpha^{2}_{1}/\alpha^{2}_{0}$. Then we can cover $\mathbb{CP}^{1}\times\mathbb{CP}^{1}$ by four charts
\begin{equation}
\mathbb{X}_{\Delta_{0}}=\{(z_{1},z_{2})\},
~\mathbb{X}_{\Delta_{1}}=\{(z^{-1}_{1},z_{2})\},~
\mathbb{X}_{\Delta_{2}}=\{(z_{1},z^{-1}_{2})\},~
\mathbb{X}_{\Delta_{3}}=\{(z^{-1}_{1},z^{-1}_{2})\},
\end{equation}
where $\check{\Delta}_{0}=\Delta_{0},\check{\Delta}_{1}=\Delta_{1},\check{\Delta}_{2}=\Delta_{2}$, and $\check{\Delta}_{3}=\Delta_{3}$ are shown in Figure 1a.
 The fan $\Sigma$ for $\mathbb{CP}^{1}\times\mathbb{CP}^{1}$ has edges spanned by $(1,0),(0,1),(-1,0),(0,-1)$. This gives the toric variety $\mathbb{X}_{\Sigma}=\mathrm{Specm}\mathbb{C}[S_{\Sigma}]$. We can also  obtain following invariant toric subvarieties,
 \begin{equation}\{0\}\times\mathbb{CP}^{1},~\{\infty\}\times\mathbb{CP}^{1},
 ~\mathbb{CP}^{1}\times\{0\},~\text{and}~\mathbb{CP}^{1}\times\{\infty\},
\end{equation}
and four intersection points as also shown in Figure 1b. The projection onto either coordinate defines a morphism of fans from $\Delta$ to the standard $\mathbb{CP}^{1}$. The corresponding morphisms of toric varieties are just the two projections onto the respective $\mathbb{CP}^{1}$ factors. Now,  we can constructed a projective toric variety
by mapping
$(\mathbb{C}^{\times})^{2}=\mathbb{X}_{\Sigma}\longrightarrow\mathbb{CP}^{3}$ define  by
\begin{equation}
(z_{1},z_{2})\longmapsto Z=(1,z_{1},z_{2},z_{1}z_{2})=(\alpha^{1}_{0}\alpha^{2}_{0},\alpha^{1}_{1}\alpha^{2}_{0},
\alpha^{1}_{0}\alpha^{2}_{1},\alpha^{1}_{1}\alpha^{2}_{1}).
\end{equation}
The coordinate ring of the projective toric variety is gives by
\begin{equation}
R_{\Sigma}=\mathbb{C}[S_{\Sigma}]\cong \mathbb{C}[\alpha_{00},\alpha_{01},\alpha_{10},\alpha_{11}]/
\langle\alpha_{00}\alpha_{11}-\alpha_{01}\alpha_{10}\rangle
\end{equation}
and by definition we have $\mathbb{X}_{\Sigma}=\mathrm{Specm} R_{\Sigma}$.
\begin{figure}
  \includegraphics[width=9cm,height=3cm,keepaspectratio,clip]{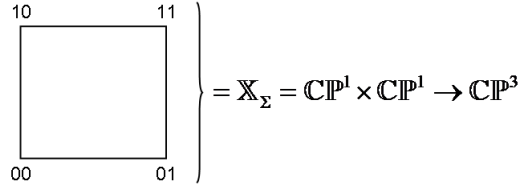}\\
  \caption{The projective toric variety   that represents the space of separable set of a pair of qubits. }\label{Figure 2}
\end{figure}
This construction is also shown in Figure 2.  Now, if we set $\lambda=1$ in equation (\ref{mono}), then the monic Laurent monomials form a multiplicative group and we can construct toric varieties by the mapping $\varphi:\mathbb{Z}^{n}\longrightarrow\mathbb{C}[z,z^{-1}]$ defined by $\beta\longmapsto z^{\beta}$ which provides an isomorphism between the additive group $\mathbb{Z}^{n}$ and the multiplicative group of moinic Larurent Monomials. For example
\begin{equation}01\equiv(0,1)\longrightarrow z^{0}_{2}z^{1}_{1}=z_{1}=\alpha^{1}_{1}/\alpha^{1}_{0}.
\end{equation}
 Note that the projective toric variety $\mathbb{X}_{\Sigma}=\mathrm{Specm} R_{\Sigma}$ is the space of  separable state for pure two qubit state and we  can construct a measure of entanglement for this state by taking the norm of polynomials that generate the ideal of this projective toric variety. Such measure entanglement is called concurrence \cite{Hosh1}. We will also see such  a construction for three-qubit states.

Next, we will discuss a three-qubit state $\ket{\Psi}=\sum^{1}_{k_{1},k_{2},k_{3}=0}
\alpha_{k_{1}k_{2}k_{3}} \ket{k_{1}k_{2}k_{3}}$. For this  state the space of separable state is given by the Segre embedding of $\mathbb{CP}^{1}\times\mathbb{CP}^{1}\times\mathbb{CP}^{1}=
\{((\alpha^{1}_{0},\alpha^{1}_{1}),(\alpha^{2}_{0},\alpha^{2}_{1}),(\alpha^{3}_{0},\alpha^{3}_{1}))): (\alpha^{1}_{0},\alpha^{1}_{1})\neq0,~(\alpha^{2}_{0},\alpha^{2}_{1})\neq0
,~(\alpha^{3}_{0},\alpha^{3}_{1})\neq0\}$. Let $S=\mathbb{Z}^{3}$ and consider the polytope $\Delta$ centered at the origin with vertices $(\pm1,\pm1,\pm1)$. This gives the toric variety $\mathbb{X}_{\Delta}=\mathrm{Specm}\mathbb{C}[S_{\Delta}]$. To describe the fan of $\mathbb{X}_{\Delta}$, we observe that the polar $\Delta^{\circ}$ is the octahedron with vertices $\pm e_1,\pm e_2, \pm e_3$. Thus the normal fan is formed from the faces of the octahedron which gives a fan $\Sigma$ whose 3-dimensional cones are octants of $\mathbb{R}^{3}$. Thus this shows that the toric variety $\mathbb{X}_{\Sigma}=\mathbb{CP}^{1}\times\mathbb{CP}^{1}\times\mathbb{CP}^{1}$.
Now, for example, let $z_{1}=\alpha^{1}_{1}/\alpha^{1}_{0}$,
 $z_{2}=\alpha^{2}_{1}/\alpha^{2}_{0}$, and $z_{3}=\alpha^{3}_{1}/\alpha^{3}_{0}$. Then we have mapping
$(\mathbb{C}^{\times})^{3}\longrightarrow\mathbb{CP}^{7}$ given by
\begin{equation}
(z_{1},z_{2},z_{2})\longmapsto Z=(1,z_{1},z_{2},z_{3},z_{1}z_{2},z_{1}z_{3},z_{2}z_{3},z_{1}z_{2}z_{3}),
\end{equation}
where $(Z=
(\alpha^{1}_{0}\alpha^{2}_{0}\alpha^{3}_{0},\alpha^{1}_{1}\alpha^{2}_{0}\alpha^{3}_{0},
\ldots,\alpha^{1}_{1}\alpha^{2}_{1}\alpha^{3}_{1})$.
This construction is also shown in Figure 3, where for example
\begin{equation}011\equiv(0,1,1)\longrightarrow z^{0}_{1} z^{1}_{2} z^{1}_{3}=z_{2}z_{3}=\alpha^{2}_{1}/ \alpha^{2}_{0} \cdot \alpha^{3}_{1}/\alpha^{3}_{0}.
\end{equation}
\begin{figure}\label{fig3e}
  \includegraphics[width=11cm,height=4cm,keepaspectratio,clip]{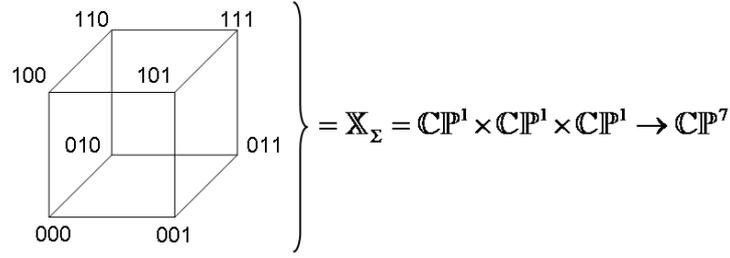}\\
  \caption{The projective toric variety that represents the space of separable set of a pure three-qubit state.}\label{Figure 3}
\end{figure}
 Moreover the coordinate ring of this toric variety is given by
\begin{equation}
\mathbb{C}[S_{\Sigma}]\cong \mathbb{C}[\alpha_{000},\alpha_{001},\ldots,\alpha_{111}]/
\langle g_{1},g_{2},\ldots, g_{12}\rangle,
\end{equation}
where $\mathrm{g}_{1}=\alpha_{000}\alpha_{110}-\alpha_{010}\alpha_{100}$,
$\mathrm{g}_{2}=\alpha_{001}\alpha_{111}-\alpha_{011}\alpha_{101}$,
$\mathrm{g}_{3}=\alpha_{001}\alpha_{101}-\alpha_{001}\alpha_{100}$,
$\mathrm{g}_{4}=\alpha_{010}\alpha_{111}-\alpha_{011}\alpha_{110}$,
$\mathrm{g}_{5}=\alpha_{000}\alpha_{011}-\alpha_{001}\alpha_{010}$,
$\mathrm{g}_{6}=\alpha_{100}\alpha_{111}-\alpha_{101}\alpha_{110}$,
$\mathrm{g}_{7}=\alpha_{000}\alpha_{111}-\alpha_{001}\alpha_{110}$,
$\mathrm{g}_{8}=\alpha_{000}\alpha_{111}-\alpha_{010}\alpha_{101}$,
$\mathrm{g}_{9}=\alpha_{000}\alpha_{111}-\alpha_{011}\alpha_{100}$,
$\mathrm{g}_{10}=\alpha_{001}\alpha_{110}-\alpha_{010}\alpha_{101}$,
$\mathrm{g}_{11}=\alpha_{010}\alpha_{101}-\alpha_{011}\alpha_{100}$,
and
$\mathrm{g}_{12}=\alpha_{010}\alpha_{101}-\alpha_{011}\alpha_{100}$. Note also that the coordinate ring  $\mathbb{C}[S_{\Sigma}]$ of  the projective toric variety $\mathbb{X}_{\Sigma}=\mathrm{Specm}\mathbb{C}[S_{\Sigma}]$ is the space of  separable state for pure three-qubit state and we can construct a measure of entanglement for this state by taking the norm of polynomial, with some multiplicity, that generate the ideal of this projective toric variety. Such measure of entanglement is also called concurrence \cite{Hosh1}.

\begin{figure}\label{fig4}
  \includegraphics[width=10cm,height=4cm,keepaspectratio,clip]{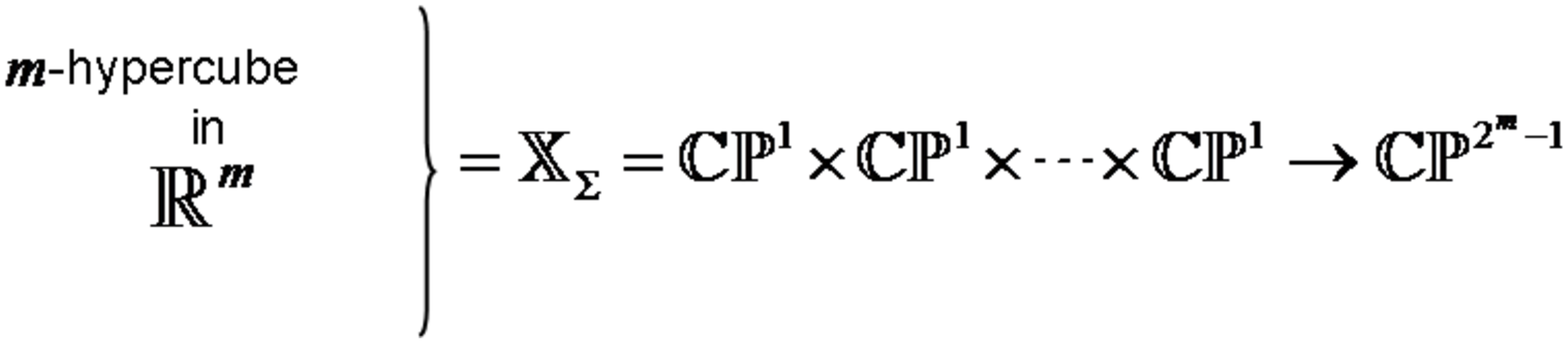}\\
  \caption{The projective toric variety  that represents space of  separable set of a pure multi-qubit state.}\label{Figure 4}
\end{figure}

Next, we will discuss a multi-qubit state $\ket{\Psi}=\sum^{1}_{k_{1},\ldots,k_{m}=0}
\alpha_{k_{1}\cdots k_{m}} \ket{k_{1}\cdots k_{m}}$. For this  state the separable state is given by the Segre embedding of $\mathbb{CP}^{1}\times\mathbb{CP}^{1}\times\cdots\times\mathbb{CP}^{1}=
\{((\alpha^{1}_{1},\alpha^{1}_{2}),\ldots,(\alpha^{m}_{1},\alpha^{m}_{2}))): (\alpha^{1}_{1},\alpha^{1}_{2})\neq0,~\ldots
,~(\alpha^{m}_{1},\alpha^{m}_{2})\neq0\}$. Let $S=\mathbb{Z}^{n}$ and consider the polytope $\Delta$ centered at the origin with vertices $(\pm1,\ldots,\pm1)$.  The polar $\Delta^{\circ}$ is the a polytope with vertices $\pm e_1,\ldots, \pm e_m$. Thus the normal fan is formed from the faces of this polytope which gives a fan $\Sigma$ whose $m$-dimensional cones are in $\mathbb{R}^{m}$, see also Figure \ref{fig4}. Thus this shows that the toric variety $\mathbb{X}_{\Sigma}=\mathbb{CP}^{1}\times\mathbb{CP}^{1}\times\cdots\times\mathbb{CP}^{1}$.
Now, let $z_{1}=\alpha^{1}_{1}/\alpha^{1}_{0},z_{2}=\alpha^{2}_{1}/\alpha^{2}_{0},\ldots z_{m}=\alpha^{m}_{1}/\alpha^{m}_{0}$. Then we have mapping
$(\mathbb{C}^{\times})^{m}\longrightarrow\mathbb{CP}^{2^{m-1}}$ given by
\begin{equation}
(z_{1},z_{2},\ldots, z_{m})\longmapsto Z=(1,z_{1},z_{2},\ldots,z_{m},z_{1}z_{2},\ldots,z_{1}z_{2}\cdots z_{m}),
\end{equation}
where $(Z=
(\alpha^{1}_{0}\alpha^{2}_{0}\cdots\alpha^{m}_{0},\alpha^{1}_{1}\alpha^{2}_{1}\cdots\alpha^{m}_{1},
\ldots,\alpha^{1}_{1}\alpha^{2}_{1}\cdots \alpha^{m}_{1})$.
Moreover, the coordinate ring of this projective toric variety is given by
\begin{equation}
R_{\Sigma}=\mathbb{C}[S_{\Sigma}]\cong \mathbb{C}[\alpha_{00\ldots
0},\alpha_{00\ldots
1},\ldots,\alpha_{11\ldots
1}]/\mathcal{I}(\mathcal{A})
,
\end{equation}
where $\mathcal{I}(\mathcal{A})=\langle \alpha_{k_{1}k_{2}\ldots
k_{m}}\alpha_{l_{1}l_{2}\ldots l_{m}} - \alpha_{k_{1}k_{2}\ldots
l_{j}\ldots k_{m}}\alpha_{l_{1}l_{2} \ldots
k_{j}\ldots l_{m}}\rangle_{\forall j;k_{j},l_{j}=0,1}$. Thus the coordinate ring $R_{\Sigma}$ of the projective toric variety $\mathbb{X}_{\Sigma}=\mathrm{Specm} R_{\Sigma}$ with its subvarieties describes the space of separable states in a multi-qubit quantum systems in the most simplest way.
\section{Conclusion}
To summarize, after a very long introduction and technicality we have managed to describe complex projective space of any pure quantum systems in terms of algebraic toric varieties and projective embedding of this varieties in a suitable complex projective  varieties. We have shown that complex space of any quantum system can be corresponds to  a  toric variety of  fan which was constructed by gluing together toric varieties of polytopes. Moreover, we have seen that the projective toric varieties are coordinates ring of space of separable multipartite states.  This is a very simple and systematic way of looking at structure of any quantum systems be it a single qubit state or a complex multi-qubit state. We hope that these geometrical results can contribute to our better understanding of complex quantum systems with applications in the field of quantum computing.

\begin{flushleft}
\textbf{Acknowledgments:} The  author  acknowledges the
financial support of the Japan Society for the Promotion of Science
(JSPS).
\end{flushleft}

\end{document}